\documentclass{mhd}     
\usepackage{graphicx}
\mhdhead{0}{0}{1}	% MHD header (vol., issue, first page)

\title{Numerical and theoretical framework for the DRESDYN
precession dynamo experiment}

\author{F.~Pizzi\inst{1}\inst{,2}, A.~Giesecke\inst{1}, 
J.~\v{S}imkanin\inst{3}, V.~Kumar\inst{1}, \\ 
T.~Gundrum\inst{1}, F.~Stefani\inst{1}}

\institute{Helmholtz-Zentrum Dresden-Rossendorf, Institute of Fluid Dynamics,
Bautzner Landstrasse 400, D-01328 Dresden, Germany
\and
Department of Aerodynamics and Fluid Mechanics, Brandenburg University 
of Technology, Cottbus-Senftenberg, 03046 Cottbus, Germany 
\and
Institute of Geophysics of the Czech Academy of Sciences, 
Bo{\v{c}}n{\'i} II/1401, 141 31 Praha 4 -- Spo{\v{r}}ilov, Czech Republic} 					

\begin{document}

\maketitle

\begin{abstract}%
The upcoming DRESDYN (DREsden Sodium facility for DYNnamo and
thermohydraulic studies) precession experiment will test the possibility 
to achieve magnetohydrodynamic dynamo action solely driven by precession. 
Here, after the description of the experimental facility, we present the 
results from direct numerical simulations with the aim to understand the 
flow behavior and its dynamo capability. The main conclusion is that in 
the nonlinear regime the nutation angle is an essential governing parameter 
which determines the flow structures and the possibility of dynamo action. 
We obtain clear indications about the optimum configuration for the future 
experimental runs.
\end{abstract}
      
\section*{Introduction.}
Precession is believed to be a complementary energy source for
the geodynamo \cite{malkus1968} or the ancient lunar dynamo \cite{dwyer2011}. 
Several laboratory experiments \cite{manasseh1992,meunier2008,herault2015} 
and numerical simulations \cite{nore2011,kong2015,giesecke2018,giesecke2019} 
have challenged the corresponding flow problem in cylindrical geometry (which 
shows commonalities with spheroidal topology) in order to understand the 
basic hydrodynamic mechanisms. One of the main motivations in selecting 
cylindrical geometry to achieve a precession driven dynamo was the experimental 
work by Gans \cite{gans1971} who had observed an amplification of an imposed 
magnetic field by factor 3.\\
The DRESDYN precession experiment at HZDR (Helmholtz-Zentrum Dresden-Rossendorf) 
consists of a cylinder with radius 1 m and height 2 m which will rotate at up 
to 10 Hz and precess at up to 1 Hz (Fig.~1) \cite{stefani2012,stefani2015}. 
The experimental studies so far were performed in a down-scaled cylinidrical 
container (R=H/2=0.163 m) filled with water in order to provide detailed 
measurements of the flow structure \cite{herault2015}. At the same time 
non-linear hydrodynamic simulations were carried out and compared with 
the experimental results \cite{giesecke2018}. In this paper we present 
results from DNS (direct numerical simulations) performed at several 
precession angles in order to find the dynamical characteristics to be 
expected in the future experiments. These hydrodynamic simulations provide 
the flow structures which are used as input for kinematic dynamo simulations 
with the goal to assess in which parameter ranges the occurrence of dynamo 
action is most likely.\\
\section{The DRESDYN precession experiment.}
\begin{figure}[h!]
\centering
\includegraphics[width=0.99\textwidth]{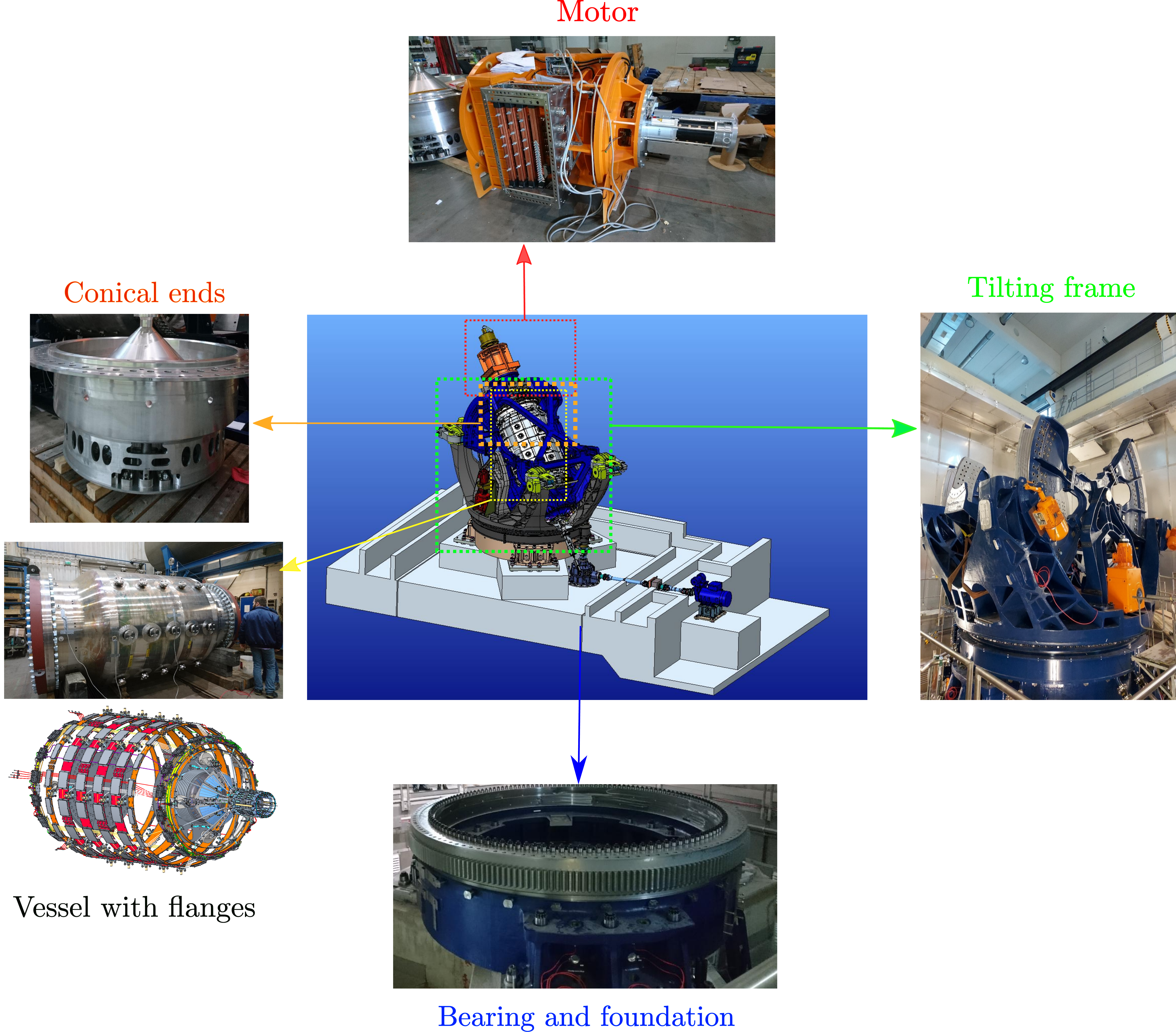}
\caption{DRESDYN precession experiment facility with all main components.
\label{fig:dresdyn_parts}}
\end{figure}

As shown in Fig.~\ref{fig:dresdyn_parts} the precession dynamo experiment 
is a large and complex machine which operates at the edge of technical 
feasibility. The ``core" of this facility is represented by a cylindrical 
stainless-steel vessel whose aspect ratio ($\Gamma=2$) is the most efficient 
one to drive an intense flow. On the sidewall 40 flanges are mounted which 
will contain the probes to measure the flow field pressure in the preliminary 
water experiments. They will be complemented by Hall sensors in the later 
sodium experiment. The vessel is fixed in a traverse and pylons which 
jointly represent the tilting frame which makes it possible to vary the 
nutation angle of the cylinder. During the dynamo experiments, the container 
contains 8 tons of liquid sodium. Because of the high flammability of the 
working fluid a fire extinguishing system with liquid argon has been installed. 
The mechanical power required to rotate the cylinder is provided by an electric 
motor of 900 kW while for the rotating turntable it is 540 kW. Due to the 
extremely high gyroscopic torque (up to 8 MNm) the basement has been built 
with ferro-concrete material and the foundation sits on 7 pillars with a 
depth of 22 m. The machine will allow to achieve a Reynolds number of 
$Re \approx 10^{8}$ and a magnetic Reynolds number of $Rm \approx 700$ 
making it possible to investigate hydrodynamic and magnetohydrodynamic 
ranges so far not reached. 

\section{Mathematical formulations and numerical methods.}
\subsection{Hydrodynamic model.}
We consider an incompressible fluid enclosed in a cylindrical container 
which is subject to precession. The main geometrical parameter is the 
aspect ratio $\Gamma$, i.e. the ratio of height over radius, which is 
fixed to 2. The nutation angle $\alpha$, whose influence is the main focus in 
this paper, is measured from the precession to the rotation axes (Fig. 1). 
The governing equations read \cite{pizzi2021b}
\begin{eqnarray}\label{eq:dimensionless_ns}
\frac{\partial {\bf{u}}}{\partial t} + {\bf{u}} \cdot {\bf{\nabla u}}   
= - {\bf{\nabla}}P + \frac{1}{Re} {\bf{\nabla}}^{2} {\bf{u}} - 2 {\bf{\Omega}} 
\times {\bf{u}} + \frac{d {\bf{\Omega}}}{dt} \times {\bf{r}} \:, 
\quad \nabla \cdot \bf{u}=0 \: , 
\end{eqnarray}
with $\bf{u}$ being the flow velocity, P the reduced pressure, 
${\bf{\Omega}}_{c}$ the rotation angular velocity and ${\bf{\Omega}}_{p}$ 
the precession angular velocity. The other governing parameters are the 
Reynolds and Poincar\'{e} numbers:

\begin{eqnarray}\label{eq:groups_dimensionles} 
Re = \frac{R^{2}\left| \Omega_{c}+\Omega_{p}\cos \alpha\right|}{\nu}, 
\quad \pm Po=\frac{\Omega_{p}}{\Omega_{c}} \: ,
\end{eqnarray}
where the $\pm$ in front of the precession ratio means prograde (the 
projection of the turntable rotation on the cylinder rotation is positive) 
or retrograde motion (projection is negative). In order to postprocess the 
results from simulations we decompose the fluid flow into its inertial 
mode components \cite{kong2015}

\begin{eqnarray}
\label{eq:decomp}
{\bf{u}}(\bf{r}, t)   &  =  & 
\sum_{m=0}^{M}\sum_{k=0}^{K} \sum_{n=0}^{2N} 
\frac{1}{2} \left[ A_{mkn}(t) \: {\bf{u}}_{mkn}\left(z,r, \varphi \right) 
+ c.c \right] +  
\bf{\widetilde{u}}^{\rm{bl}}(\bf{r})
\nonumber
%}
\end{eqnarray}
where $c.c$ means complex conjugate, $\bf{\widetilde{u}}^{bl}$ is 
the boundary layer velocity and $\bf{r}$ is the position vector. 
The triplet $(m,k,n)$ represents the azimuthal, axial and radial wave 
numbers. The amplitude of each inertial mode $A_{mkn}$ is computed 
projecting the simulated velocity on $u_{mkn}$ which are the eigenfunctions 
of the linearized inviscid form of Eq.~\ref{eq:dimensionless_ns} \cite{greenspan1968}.

\begin{figure}[h!]
\centering
\includegraphics[width=0.95\textwidth]{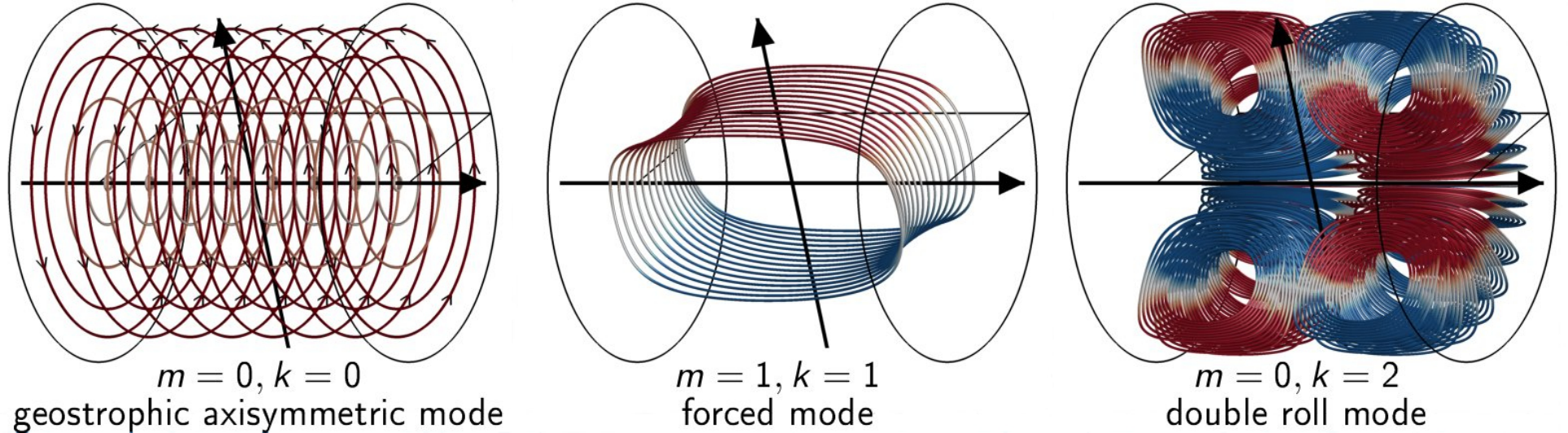}
\caption{Three main flow structures emerging in a fluid-filled precessing cylinder.
\label{fig:3_structures}}
\end{figure}

Figure \ref{fig:3_structures} represents the 3D structures of the most 
prominent inertial modes occurring in a fluid filled precessing cylinder: 
the central one is the directly forced mode caused by the gyroscopic 
effect on the fluid flow which dominates in the laminar regime, i.e 
for weak forcing. Once the precession ratio increases, the other two 
modes emerge due to the nonlinear phenomena enriching the dynamics 
of the flow. In particular the (0,0) mode resembles a columnar vortex 
which counteracts the solid body rotation while the (0,2) is a 
poloidal flow analogous to the Taylor vortices found in other 
fluid dynamics problem \cite{giesecke2018,pizzi2021b}.

\subsection{MHD model: Induction equation.}
We run purely hydrodynamic simulations until the flow
field achieves a statistically stable regime. Then, we insert the 
time-averaged velocity $\langle \bf{u} \rangle$ into the induction equation 
\begin{eqnarray}\label{eq:induction}
\frac{\partial {\bf{B}}}{\partial t} = {\bf{\nabla}} \times 
\left( \langle {\bf{u}} \rangle \times {\bf{B}} - \frac{1}{Rm}\nabla 
\times {\bf{B}} \right) \: , \quad {\bf{\nabla}} \cdot {\bf{B}}=0 \: .
\end{eqnarray}
The second condition expressed in Eq.~(\ref{eq:induction}) represents 
the solenoidal nature of the magnetic field. For the time-averaged 
velocity field $\langle {\bf{u}} \rangle$ considered constant, the 
solution of the linear evolution equation has the form 
${\bf{B}}={\bf{B}}_{0} \exp \left( \sigma \: t\right)$ with $\sigma$ 
representing the eigenvalue. In Eq.~(3) $Rm$ is the magnetic Reynolds 
number, defined as $Rm=Re \: \nu/\eta$, where $\eta$ is the magnetic 
diffusivity of the liquid metal. While for the hydrodynamic problem 
we use a spectral element-Fourier code \cite{semtex2019} with no 
slip boundary condition, we solve the induction equation through 
a finite volume scheme with constraint transport in order to 
ensure the divergence-free nature of the magnetic field. As 
boundary conditions we use the so-called \textit{pseudo vacuum} 
condition i.e, only the tangential components of the magnetic 
fields vanish at the wall. Even if this model lack of physical 
consistency it is a good approximation since it is computationally 
faster and the results are not far from those using correct 
boundary conditions \cite{giesecke2012}. We present results 
at fixed $Re=6500$ for which we study the role of the nutation angle 
which will be shown to be substantial both for hydrodynamic and 
magnetohydrodynamic behavior.
\section{Hydrodynamic Results.}
\begin{figure}[h!]
\centering
\includegraphics[width=0.95\textwidth]{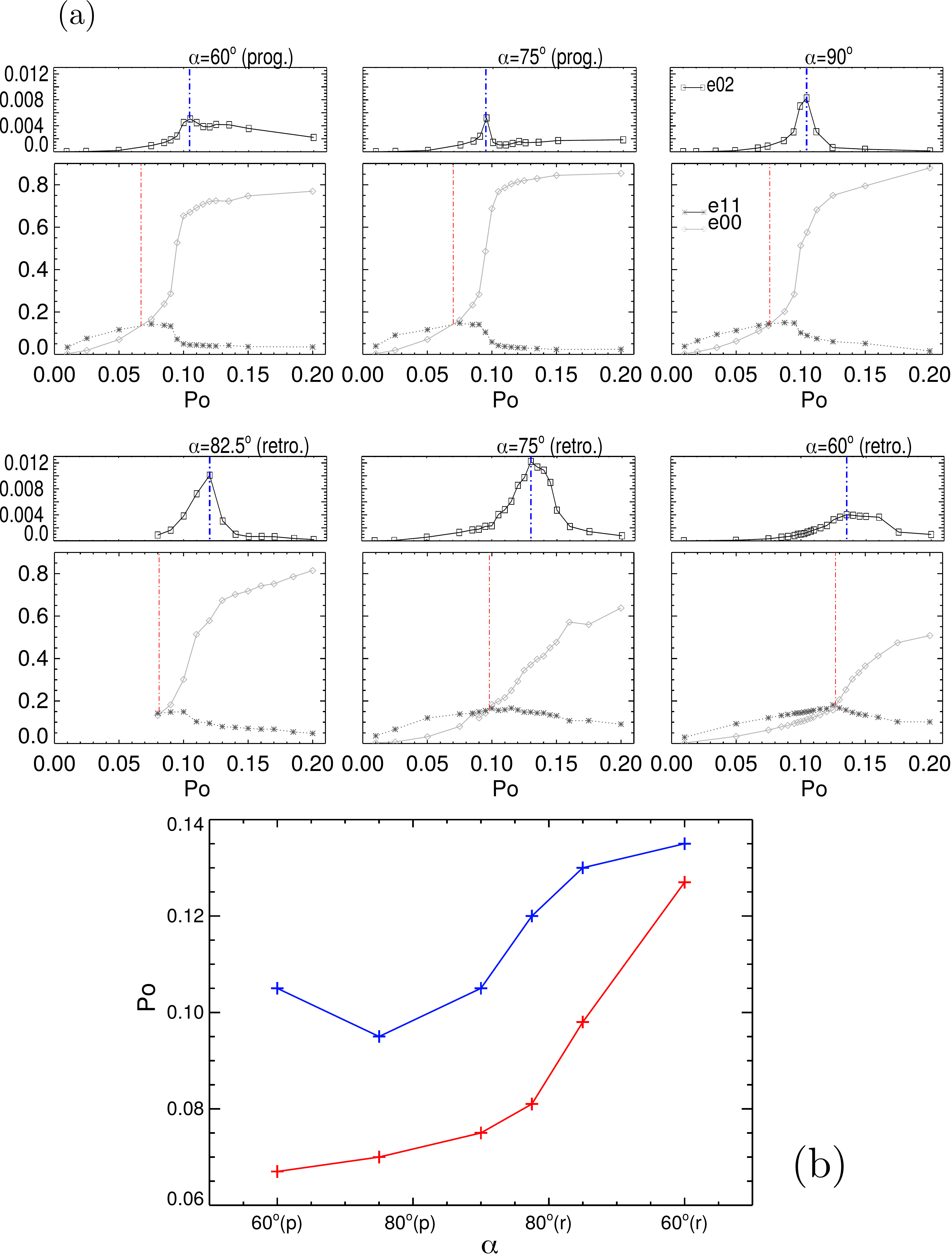}
\caption{(a) Energy of the three main inertial modes in a 
precessing cylinder, as a  function of the precession ratio 
for 6 different nutation angles. The larger plots show 
$e_{00}$ and $e_{11}$ while the smaller plots show $e_{02}$. 
The red dotted lines mark the energetic crossing between the 
two modes, i.e the passage from a directly forced dominated 
to a geostrophic dominated bulk flow. The blue dotted lines 
show the peak of $e_{02}$. (b) The corresponding precession 
ratios for peak and crossing versus the angles. Here $(p)$ 
and $(r)$ mean respectively prograde and retrograde flow.
\label{fig:inertial_modes_angles}}
\end{figure}
The most dominant modes emerging in a precessing cylinder, 
schematically shown in Fig. \ref{fig:3_structures}, are quantified 
in terms of their kinetic energies (normalized with the energy of 
solid body rotation) in Fig.~\ref{fig:inertial_modes_angles}(a). 
For each angle we plot the axisymmetric-geostrophic energy $e_{00}$, 
the directly forced flow energy $e_{11}$ and (in a separated 
panel due to the difference in order of magnitude) the axisymmetric 
poloidal energy $e_{02}$. The prograde and $90^{\circ}$ cases 
(top row) show a clear transition between a weak forcing region 
dominated by $e_{11}$ and a strong forcing region where $e_{00}$ 
accounts for $ \approx 80\%$ of the total energy. For the retrograde 
cases, only the $82.5^{\circ}$ (for which we do not have simulation 
for $Po<0.075$) case undergoes a marked breakdown of $e_{11}$ 
with a ``jump'' of $e_{00}$ while the smaller $\alpha$ show a 
smoother profiles of the $e_{11}$ when $Po$ increases. The red 
curves mark the crossing of the two energy profiles which occur 
at different $Po$ as plotted as a red solid line 
in \ref{fig:3_structures}(b). With these points coincides a shift 
from the directly forced dominated to a geostrophic dominated 
flow which has an impact also on the boundary 
layers \cite{pizzi2021a}. The last contribution shown is $e_{02}$. 
Also for this component the nutation angle plays a crucial role 
in terms of magnitude, shape of the profiles and the location of 
the peaks (dotted blue curve). Indeed we plot the position 
of max($e_{02}$) as a function of precession ratio in  
Fig.~\ref{fig:inertial_modes_angles}(b) observing a shift 
towards larger $Po$ for the retrograde cases.  
\section{Dynamo action}
\begin{figure}[h!]
\centering
\includegraphics[width=0.95\textwidth]{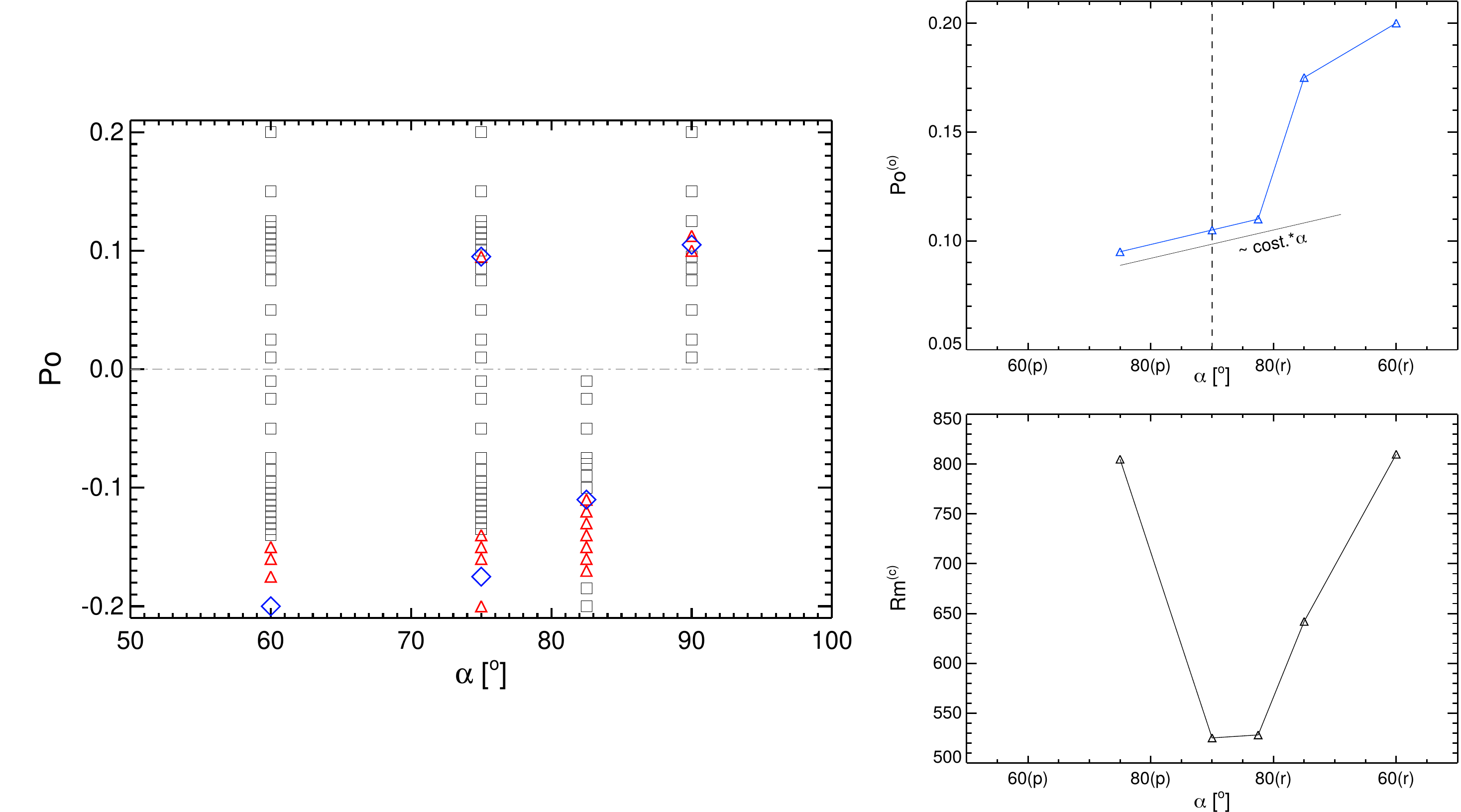}
\caption{Influence of the nutation angle $\alpha$ on the dynamo 
action for a fixed Reynolds number $Re=6500$. Left: phase 
diagram. Black symbols are the no-dynamo cases; red symbols 
are dynamo cases and blue diamonds denote the largest growth 
rate of the magnetic energy, i.e the optimum for dynamo. 
Right top: optimum precession ratio $Po^{o}$ in terms of 
dynamo action versus the nutation angle. Right bottom: 
critical magnetic Reynolds number  versus the nutation angle.
\label{fig:alpha_dependence}}
\end{figure}
Having observed that the nutation angle (and of course the 
precession ratio Po) influences substantially the flow field, we 
check now the role of $\alpha$ for the precession driven flow 
to act as dynamo, i.e. to have a positive growth rate of the 
magnetic energy. The first general result is the phase 
diagram (Po, $\alpha$) shown in Fig.~\ref{fig:alpha_dependence} 
where the red symbols show the dynamo cases, while the black 
symbols stands for no-dynamo. We observe that the retrograde 
motion (negative semiplane) is more prone to dynamo action while 
the prograde cases show only few red points. Interestingly 
the $\alpha=60^{\circ}$ case is not able to drive a dynamo at 
all. By contrast, notice the more extended region of red symbols 
at $\alpha=82.5$ retrograde. The blue diamonds mark the 
strongest dynamo for each angle and they are plotted in the 
top-right figure as optimum precession ratio $Po^{o}$ function 
of $\alpha$. Notably the profile presents an asymmetry with 
respect to $\alpha$ since the linear trend includes also the 
retrograde case $\alpha=82.5^{\circ}$ and the jump does not 
coincide with $\alpha=90^{\circ}$. In the right bottom plot 
is shown the critical magnetic Reynolds number (i.e. the 
lowest $Rm$ for the occurrence of dynamo action) versus the 
nutation angle. Evidently, the role of $\alpha$ is crucial 
in order to reduce as much as possible the threshold for 
the onset of dynamo, and it is clear that the optimum range 
is $90^{\circ} < \alpha< 82.5^{\circ}$ retrograde.\\ 
\begin{figure}[h!]
\centering
\includegraphics[width=0.95\textwidth]{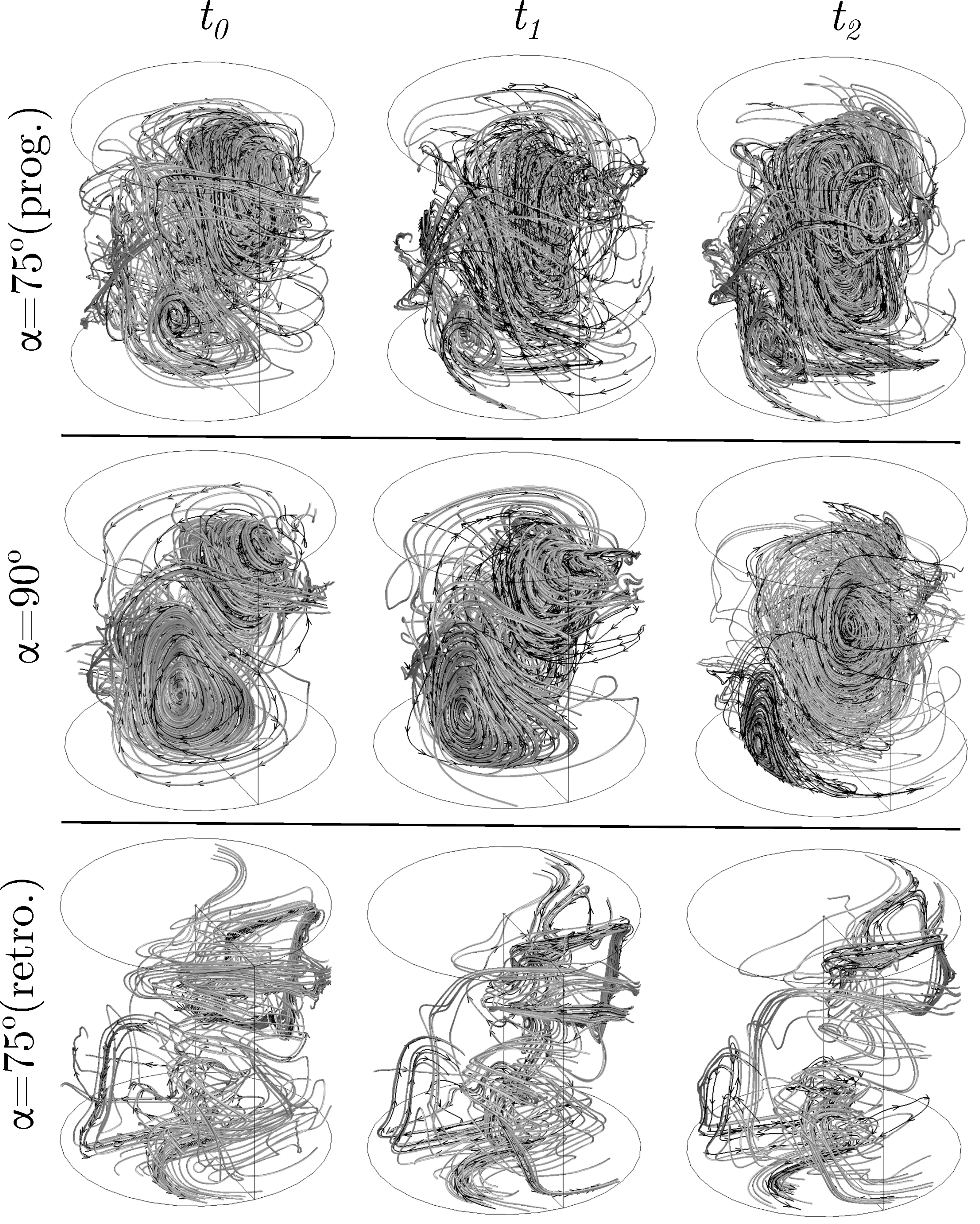}
\caption{Influence of the nutation angle $\alpha$ on the 
time-evolution of the three dimensional magnetic field 
inside the precessing cylinder. Columns represent three 
different times (respectively 1,1.25,and 1.5 of the 
dimensionless diffusion time $1/Rm$) while the rows 
represent three different angles. Top row is for 
$Rm=850$, middle row $Rm=550$ and bottom row $Rm=700$. 
\label{fig:3d_mf}}
\end{figure}
As a final result we show in Fig.~\ref{fig:3d_mf} the 
three dimensional magnetic field lines developed inside 
the container at three different times (columns) for 
three different nutation angles. The main outcome is 
that $\alpha$ causes a substantial difference in the 
geometry and the time-evolution of the field. The 
$\alpha=75^{\circ}$ retrograde case presents less 
dense field lines with quite chaotic shape, while 
the $\alpha=90^{\circ}$ shows the most coherent 
topology characterized by two clear vortices with 
opposite rotations (visible as black arrows) which 
change over time: e.g. at $t_{0}$ the bottom large 
vortex is clockwise while at $t_{1}$ it becomes counterclockwise. 

\section{Conclusions and outlook}
The nutation angle is a crucial governing parameter for 
precession driven flows in cylindrical geometry  whose role 
is far from trivial, especially in the nonlinear regime. 
It determines the flow topology and the magnitude of 
particular structures both for the flow and the dynamo 
generated magnetic field. As a consequence the possibility 
to achieve dynamo action is strongly influenced by the 
angle. The results of this paper give general indications 
about the best region in the phase space for the dynamo 
action: the optimum angle was found to be around 
$90^{\circ} < \alpha < 82.5^{\circ}$ retrograde. Future 
studies will focus on the extension of the analysis on 
other Reynolds number and to the impact of realistic 
magnetic boundary conditions.

\Thanks{This project has received funding from the European 
Research Council (ERC) under the European Union's Horizon 2020 
research and innovation program (Grant Agreement No. 787544), and from 
Deutsche Forschungsgemeinschaft under Project No. GI 1405/1-1.}

% -------------------- End of References

\lastpageno	

\end{document}